# Shall androids dream of genocides? How generative AI can change the future of memorialization of mass atrocities

**Mykola Makhortykh\*, Eve M. Zucker\*\*, David J. Simon\*\*\*, Daniel Bultmann\*\*\*\*, Roberto Ulloa\*\*\*\*\***

\*Institute of Communication and Media Studies, University of Bern
\*\*Department of Anthropology, Yale University/Weatherhead Institute of East Asian Studies, Columbia University
\*\*\*Jackson School of Global Affairs, Yale University
\*\*\*\*Humboldt-Universität zu Berlin
\*\*\*\*\*University of Konstanz/GESIS – Leibniz-Institut für Sozialwissenschaften

**Corresponding author:** Mykola Makhortykh, Institute of Communication and Media Studies, University of Bern, Switzerland, Fabrikstrasse 8, 3012 Bern, mykola.makhortykh@unibe.ch.

**ORCID IDs:** *Mykola Makhortykh (0000-0001-7143-5317), Eve M. Zucker (0000-0001-7054-0342), David J. Simon (0000-0003-0565-0718), Daniel Bultmann (0000-0001-5465-2010), Roberto Ulloa (0000-0002-9870-5505).*

**Abstract:** The memorialization of mass atrocities such as war crimes and genocides facilitates the remembrance of past suffering, honors those who resisted the perpetrators, and helps prevent the distortion of historical facts. Digital technologies have transformed memorialization practices by enabling less top-down and more creative approaches to remember mass atrocities. At the same time, they may also facilitate the spread of denialism and distortion, attempt to justify past crimes and attack the dignity of victims. The emergence of generative forms of artificial intelligence (AI), which produce textual and visual content, has the potential to revolutionize the field of memorialization even further. AI can identify patterns in training data to create new narratives for representing and interpreting mass atrocities – and do so in a fraction of the time it takes for humans. The use of generative AI in this context raises numerous questions: For example, can the paucity of training data on mass atrocities distort how AI interprets some atrocity-related inquiries? How important is the ability to differentiate between human- and AI-made content concerning mass atrocities? Can AI-made content be used to promote false information concerning atrocities? This article addresses these and other questions by examining the opportunities and risks associated with using generative AIs for memorializing mass atrocities. It also discusses recommendations for AIs integration in memorialization practices to steer the use of these technologies toward a more ethical and sustainable direction.

## Author Note

This is the preprint version of the "Shall androids dream of genocides? How generative AI can change the future of memorialization of mass atrocities" article.

**Declarations**

*Funding*: The research has been supported by the Alfred Landecker Foundation, which provided financial support for the research time of Mykola Makhortykh, who contributed to the article as part of his project titled "Algorithmic turn in Holocaust memory transmission".

*Conflicts of interest/Competing interests*: Not applicable.

*Data availability*: Not applicable.

*Code availability*: Not applicable.

*Authors' contributions*:
- Conceptualization: Eve M. Zucker, Mykola Makhortykh, Roberto Ulloa.
- Writing - original draft preparation: Mykola Makhortykh, Eve M. Zucker, Roberto Ulloa. Writing - review and editing: Eve M. Zucker, Mykola Makhortykh, David J. Simon, Roberto Ulloa, Daniel Bultmann.
- Literature search: Mykola Makhortykh, Eve M. Zucker, Roberto Ulloa, David J. Simon. Funding acquisition: Mykola Makhortykh.
- Supervision: Mykola Makhortykh, Eve M. Zucker.

# Shall androids dream of genocides? How generative AI can change the future of memorialization of mass atrocities

**Abstract**: The memorialization of mass atrocities such as war crimes and genocides facilitates the remembrance of past suffering, honors those who resisted the perpetrators, and helps prevent the distortion of historical facts. Digital technologies have transformed memorialization practices by enabling less top-down and more creative approaches to remember mass atrocities. At the same time, they may also facilitate the spread of denialism and distortion, attempt to justify past crimes and attack the dignity of victims. The emergence of generative forms of artificial intelligence (AI), which produce textual and visual content, has the potential to revolutionize the field of memorialization even further. AI can identify patterns in training data to create new narratives for representing and interpreting mass atrocities – and do so in a fraction of the time it takes for humans. The use of generative AI in this context raises numerous questions: For example, can the paucity of training data on mass atrocities distort how AI interprets some atrocity-related inquiries? How important is the ability to differentiate between human- and AI-made content concerning mass atrocities? Can AI-made content be used to promote false information concerning atrocities? This article addresses these and other questions by examining the opportunities and risks associated with using generative AIs for memorializing mass atrocities. It also discusses recommendations for AIs integration in memorialization practices to steer the use of these technologies toward a more ethical and sustainable direction.

**Introduction**

Memorialization of mass atrocities, such as war crimes and genocides, has a special meaning to humankind. Memorialization involves the practice of preserving the memory of those who suffered, the experiences of what they endured, and the mass atrocity event itself. As such, the memorialization of collective suffering addresses many ethical and moral obligations that may be specific to a particular culture and society and reflect ideas drawn from transnational human rights and transitional justice. These obligations range from respecting and tending to the dead [66,68] and honoring those who fought against or resisted the perpetrators [31,42][1] to countering the distortion of historical facts and preventing the repetition of atrocities [17]. The construction of social memory and memorialization are thus integral to the moral reconstitution and social repair of societies in the aftermath of atrocities [18,100]. However, these obligations are not easy to fulfill. Challenges include contestation over individual and collective memories with the imperative of accounting for various perspectives of the past, but without legitimizing patently false ones [68]; balancing individual narratives versus collective narratives and local versus national and international ones [44]; temporal, political, and situational influences that may influence different narrative constructions among individuals or groups [81,101]; limited access to human and financial resources required for commemoration [68]; and the need to resist efforts to instrumentalize memorialization efforts, which subordinate social needs to short-term political gains [47].

---

[1] Khan's work focuses on rescuers in genocide and other conflict situations with the aim of highlighting their moral choices as a model for younger generations. Similarly, institutions such as Yad Vashem recognize the noble acts of those who took great risks to help and save Jewish people during the Holocaust [93]. Gruner's work, on the other hand, details the resistance by Jewish people to Nazi policies and persecution. All three cases, in different ways, honor those who resisted and worked to save those in dire situations.

As with other societal practices, digital technologies have transformed the memorialization of mass atrocities [4,20,62,63,90,92,102]. While mass media had a major influence on memorialization already in the analog era [91], the rise of digital platforms caused a profound disruption in memory practices. The unprecedented amount of digitized and digital-born content, the increased connectivity between consumers and producers of memories [35], and the fluidity between various digital formats, including images, text, and video [39] have created space for less top-down memory practices and greater autonomy and creativity in memorial works. These practices offer the prospect of giving voice to under-represented communities and enable possibilities for engaging with the difficult pasts that might not otherwise have been accessible via heritage institutions and archives [78]. At the same time, digital technologies create challenges for memorialization, potentially facilitating denialism, and distortion [32,55], while contributing to the trivialization of past suffering through the amplification of cynical attitudes toward it [36]. This latter phenomenon emerges from the potent combination of the unprecedented availability of atrocities-related content [10] and the ease with which such evidence of violence can be fabricated. Furthermore, the plethora of online memorial activities may obscure some memorial efforts in the noise of others or spark the banality that accompanies surface-level repetition in a manner reminiscent of Walter Benjamin's thoughts on the risks of mechanical reproduction [3].

The extensive volume of atrocities-related digital content prompts the need to help individuals navigate it. Since the 2000s, this task has been increasingly delegated to artificial intelligence (AI)-driven systems, such as search engines and recommendation systems, which filter content items in response to user input (e.g., search queries) and retrieve items that are viewed by the system as the most relevant for the users [48]. Despite the increasing use of such systems by the heritage institutions (e.g., USHMM digital collections' search; https://collections.ushmm.org/search/) and commercial platforms (e.g., Google search), the discussion of the opportunities and risks that AI systems create for the memorialization of mass atrocities remains limited (for some exceptions, see [6,43,56,89]). For example, recent research has exposed the role of different search engines in memorialization by analyzing the search results in relation to the Holocaust [59] and the Ukrainian Holodomor[2] [60] and potential distortions in the representation of mass atrocities which these systems may cause.

The situation becomes more complex with the emergence of generative AIs, which represent a profound transformation in the context of the memorialization of mass atrocities. Instead of retrieving information about atrocities and prioritizing certain content sources, generative AIs produce the textual/visual content themselves [12], as illustrated in Figure 1. The implications of this change for the memorialization of mass atrocities are substantial: we can expect not only a rapid increase of content engaging with memory about past suffering but also significant challenges in distinguishing between human- and AI-made content. The shift to generative AI also implies major changes in how information about mass atrocities is retrieved and how humans interact with such content in various contexts (e.g., education, community NGOs, international development and humanitarian entities, and government).

---

[2] Holodomor is a human-made famine that took place in the 1930s and took a particularly heavy toll on the Ukrainian population of the Soviet Union. A growing number of countries around the world recognize the Holodomor as a genocide perpetrated against the Ukrainian people.

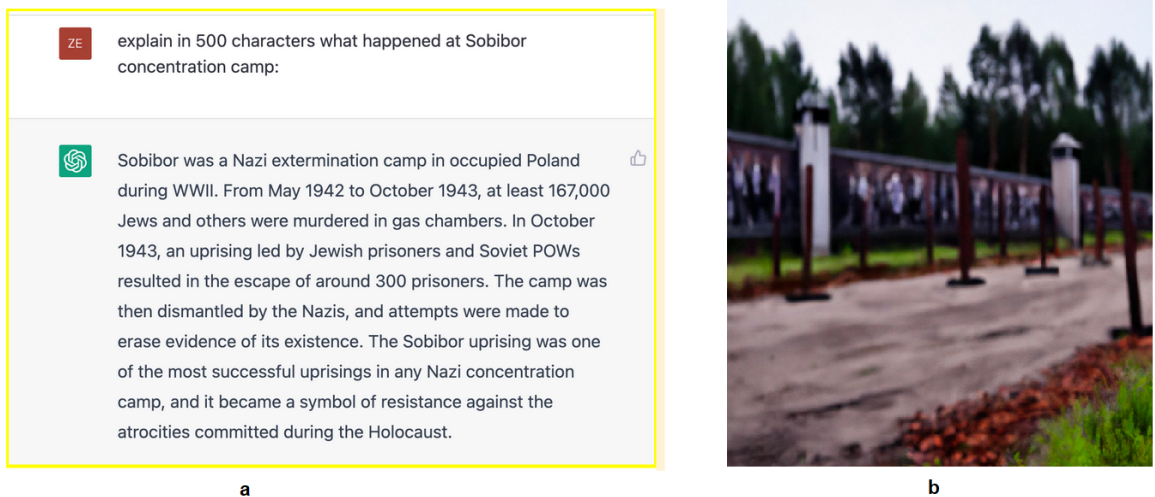

**Figure 1**. (a) ChatGPT (https://chat.openai.com/) produced in a few seconds the piece of text responding to the prompt inquiring to explain in 500 characters "what happened at Sobibor concentration camp". (b) The AI platform Rytr (https://rytr.me/), also in seconds, produced a digital rendition of an oil painting using the prompt "digital memorial of Sobibor concentration camp with representation of victims in oil paint".

The efficiency of generative AIs in producing text and illustrations can facilitate the processes of compiling and analyzing memorialization data and make efforts to memorialize the past more accessible for individuals worldwide. Yet the long-term consequences of adopting generative AIs in the context of memorializing both historical (e.g., the Holocaust) and recent atrocities (e.g., Russian war crimes in Ukraine) are uncertain. Some of the related concerns include the non-transparency of sources used to automatically generate narratives about the atrocities, the possibilities for misrepresentations of historical facts in rendered content, and the ethical problems arising from the generation of fake evidence or the manipulation of real one in a manner that changes the meaning of personal testimonies or artifacts. Additionally, it is unclear how the generated content might influence beliefs and attitudes toward a given atrocity, or how a growing distrust of some forms of digital media[3] might interact with an increasing awareness of AI capabilities and the pervasiveness of AI-rendered content in highly sensitive contexts such as the ones dealing with the memorialization of suffering.

Through our interdisciplinary perspective and experience of working in the field of memorialization, we aim to address some of these concerns and questions by reviewing the possibilities and challenges generative AI brings to the online memorialization of mass atrocities. As AI-related technologies continue to approach and exceed human capabilities in pattern recognition and processing large volumes of data in an increasing number of areas[4], it is crucial to evaluate their potential impact. Such evaluations are integral for deciding whether (and how) to incorporate generative AI in an area where technologies have to be adopted particularly carefully due to the ethical implications or whether recommending against AI use in this field is more beneficial due to its inadequacy to account for multiple ethical considerations guiding the process of atrocities' memorialization.

---

[3] On this phenomenon, see [70,71]. For instance, Newman and Fletcher [70] used survey data from nine OECD countries and found that amid growing distrust in all media, social media is the least trusted news medium.

**Possibilities of generative AI for digital memorialization of mass atrocities**

Generative AI can enable new possibilities for creating content that can be used for memorialization purposes. Such content can be either visual (motion or still) or textual and can reproduce multiple formats which are currently used in the context of digital grassroots memorialization of mass atrocities; examples of these formats range from video tributes commemorating specific instances of mass violence [55] to the Internet memes reinforcing [28] or challenging established genocide narratives [7] to the images imitating photos made at the sites of violence [52] to the drawings aiming to attract the attention of the general public to the past and present suffering [82] to the online encyclopedia entries aiming to find consensus between the different viewpoints on the past [54]. As a result, AI might induce a variety of positive developments, including expanding the realm of creation and participation in memorialization, expanding public access to knowledge about the past, helping researchers learn more about mass atrocities and the memorialization thereof, and serving as a tool to help detect and counter false narratives.

- *AI can expand the capacity to memorialize*

Generative AIs can easily produce representations of mass atrocities. The ease of access and use of these platforms enables new possibilities for expressing mourning, producing and analyzing testimony, and communicating loss – and for a small portion of the costs required to generate textual and image content to fulfill these purposes by relying on individual human efforts. These productions can then be employed on their own or combined with other digital- or analog-born materials to memorialize past and present atrocities (e.g., by populating digital memorials with AI-generated content when such content is lacking or might undermine the privacy or safety of the victims or their relatives). This is particularly true in relation to survivors engaging with an AI platform to tell their stories or as a therapist assistant [21].

Similarly, the textual descriptions of atrocities or commemorative statements generated via text-focused AIs might help connect with an international audience. For example, by enabling individuals to participate in online memorial practices in different languages without the need for translators, copyeditors, or other production assistants. Under these circumstances, image- and text-focused AIs can empower individuals who want to create, contribute to, or expand existing grassroots memorialization practices with almost limitless creativity providing their projects remain within the accepted use requirements of the platforms, for example, Midjourney's use guidelines aiming to prevent the creation of graphic or potentially harmful images through an explicit list of banned prompts [30,65].

---

[4] Notably, AI development seems well ahead of the pace of AI capability development predicted by [29, p. 729], which included "translating languages (by 2024) [and] writing high-school essays (by 2026)". See, for example, [53] on language translation and [33] on high school essay writing.

- *AI can enable new ways for the general public to learn about mass atrocities*

Engaging with generative AIs as a means to learn about mass atrocities can be a form of memorialization practice itself. Earlier, non-generative versions of AI-driven systems focused on information retrieval tasks [1]. For example, the *Let Them Think*[5] platform involved a database of Holocaust textual testimonies that could be queried for specific words and used to create lists of fragments [69]. The Shoah Foundation-produced holograms of Holocaust survivors [86,89] is a technologically similar form of AI that uses a finite database of questions with pre-recorded answers from Holocaust survivors and is employed to enable an interactive experience for individuals and groups engaging with the holograms.

A key distinction of generative AI is the ability to produce *new* narratives, store these narratives and use them as additional data for iterative training. It allows generative AIs not only to retrieve content in response to an (often limited) set of user inputs but generate content and sustain a broader range of interactions with individuals interested in exploring the past. Thus a user may have a vested interest in continuing the dialogue with the generative AI chatbot on a given topic rather than having to start at the beginning each time as in the case of more traditional conversational agents. These interactions, which occur on one-on-one user and machine bases over private accounts, can be understood as a form of memorialization in itself: for instance, when individuals engage with AIs to generate a text- or image-based narrative about a mass atrocity.

- *AI can be a tool for researchers to collect/analyze data*

The advancements in the field of AI also transform the processes for collecting and analyzing data concerning both historical and present mass atrocities. Instead of simply retrieving and ranking information sources relevant to the user queries (as non-generative AIs used to do), generative AI platforms can directly respond to user requests while also providing suggestions of information sources for further exploration (e.g., as in the case of generative AI interface for Bing search). Such functionality can potentially accelerate the process of analyzing data on mass atrocities and lead to profound transformations in how institutions provide access to these data. For example, instead of using a conventional archive search, a generative AI can be integrated into the interface to provide recommendations to the researcher in a conversational format; similarly, generative AI can be used to generate visualizations based on text descriptions to offer additional perspectives for the researcher. In addition to retrieving information in a more digestible format, generative AIs can also be used for automated content labeling that can help identify content related to mass atrocities, given AIs' capacity for this kind of task can be comparable or superior to that of humans [26,87].

There are also other capabilities of generative AI which can facilitate the analysis of data related to mass atrocities. Similar to earlier AI-driven information retrieval systems, generative AI can personalize outputs (e.g., textual content generation prompted by individual user needs and interests). For example, AI can summarize data and present several summaries emphasizing various elements to let the researcher think through the data from various perspectives. Furthermore, both text- and image-focused generative AIs can be used for creating synthetic data (see, for instance [76]), which can then be employed for training computational approaches for automated content analysis. For example, the Iraq war crimes investigation initiative UNITAD and the Germany-based non-governmental organization Mnemonic have reportedly started using AI to scan hundreds of thousands of hours of video to identify evidence of war crimes in the context of recent armed conflicts in Iraq and Syria [85].

---

[5] For a more detailed discussion of the platform and its relationship to digital memorialization see [69]. The platform itself can be found here: https://lts.fortunoff.library.yale.edu/

- *AI can be an instrument for detecting distortion and denialism*

In addition to the above-mentioned possibilities, the focus of generative AIs on interpreting and engaging with user input makes them an effective means of content classification. Broadly understood, content classification involves the assignment of specific labels to content items to detect specific attributes of such content (e.g., whether it is related to a specific subject, such as politics, or expresses a specific sentiment, for instance, positive or negative one). In the case of content related to mass atrocities, such classification can take multiple forms, including whether the statement that AI is asked to evaluate comes from a certain source or whether it might contain denialist claims. While the degree to which generative AI can perform these functions, specifically in the case of content dealing with mass atrocities, is yet to be studied, generative AI platforms, such as chatGPT, have demonstrated the ability to identify whether claims entered by the users are false in relation to diverse subjects [34] and health-related matters in particular [41], and to evaluate the degree of news outlets' credibility [95]. Under these circumstances, generative AI can potentially be used to filter out content promoting distortion or denialism of mass atrocities, with the possibility that individuals will be more open to engaging with views they find contrasting to their own if these views are expressed via a human-like conversational interface (for examples related to more conventional chatbot, see [96]).

**Threats of generative AI for digital memorialization of mass atrocities**

Despite the many advantages generative AI presents for the memorialization of atrocities, there are also risks. An amoral machine entity, AI does not assign specific meaning to the data it processes. Rather, AI is more akin to Bender and Kohler's "octopus test" [2], where a clever octopus fools people into believing it is a sentient human by communicating in the English language through statistical patterns until it fails to comprehend the implications of an unfamiliar context, or Searle's "Chinese Room" proposition [77] in which he posits he could create coherent text in Chinese by learning to manipulate symbols without actually understand the language itself. Both exercises illuminate how the capacity to impart information should not be conflated with the ability to understand the meaning – nor should it, beyond that point, be confused with the capability to make context-driven moral judgments. Yet memorialization is a realm of human thought and activity for which the abilities to make relatively fine distinctions in meaning and to exercise moral judgment are essential.

- *AI can serve as a means of enforcing hegemonic narratives and practices*

Generative AIs are trained on specific sets of data. Unless their training is diversified, AIs can take prevalent patterns in data for granted and then reiterate them in content which AIs generate. In the case of atrocities-related content, it may result in the enforcement of hegemonic narratives and representation practices, for instance, by prioritizing Western-centric views on how mass atrocities shall be remembered or interpreted and uncritically translating these views to other contexts. Such reinforcement of memory hegemonies might result in the silencing or erasure of experiences of minority groups (including the ones which were disproportionately affected by the mass atrocities) and suppression of alternative practices of memorialization [19]. It is additionally concerning in the case of authoritarian states, where hegemonic memory practices often serve as an integral means of state ideology and propaganda and where generative AIs can become an effective tool for consolidating the national memory regimes, such as in recent efforts by China [16].

The capabilities of generative AIs to enforce hegemonic narratives are further amplified by the risks of keeping users in information bubbles, which was a common concern about the non-generative AI-driven systems related to information retrieval (e.g., [9,71]). While the existence of AI-amplified information bubbles has so far found little empirical support [9] and, in some cases, such bubbles can be beneficial for society by nurturing independent thinking (e.g., [58]), the possibility of generative AIs leading their users into the "rabbit holes"[6] [8] of memory hegemonies can not be currently excluded. Further support to such concerns is provided by the observations indicating that some generative AI programs can remember the previous interactions with a particular user [97], thus contouring its answers to what the system perceives as the desired type of response the user is seeking.

Self-reinforcement or feedback loops wherein model outputs become increasingly less diverse over time [83] are another aspect of generative AIs which can potentially serve to amplify hegemonic narratives of atrocities. While the possibility of collecting user input (e.g., positive/negative evaluations of outputs of chatGPT) is important for improving the system functionality, in the case of memorialization, it might also result in prioritization of small sets of possible outputs (e.g., in terms of standardized narratives or visual representations of atrocities). Given the highly political battles over narrative creation that are often embedded in memorialization efforts (see, for instance, the case of the Holodomor [45,73] or the genocide in Rwanda [40], AI platforms – assiduously apolitical by design – might nevertheless have profound political and social consequences.

- *AI can introduce bias in analysis and make some research practices more obscure/obsolete.*

The popularity of AI can entice (or even force) researchers towards relying on AI-driven tools instead of traditional qualitative practices of studying the memorialization of mass atrocities. While it can enable new possibilities for research in the field of memorialization, it might diminish the role of the human in the loop by making obsolete interactions and forms of knowledge that require embodiment and allow for a wider range of communication beyond what the digital systems can produce or relay. Such a shift can be concerning for a multitude of reasons, including the possibility of generative AIs being subjected to certain forms of bias that, in the case of earlier AI systems, varied from the retrieval of factually incorrect information or systematic skewness of outputs (e.g., in terms of visibility of specific memorial sites and practices [59]).

---

[6] The rabbit hole denotes the process of moving from the diverse information environment to an ideologically extreme echo chamber.

- *AI can be used to facilitate and enhance censorship.*

The ability to program what AI can and cannot generate creates not only possibilities for the memorialization of mass atrocities but also risks. It might improve efforts to detect and combat denial and distortion, but it could just as easily inhibit individuals or groups in vulnerable situations, such as in authoritarian countries, from participating in memorial activities that counter the state's narrative. For instance, AIs can be prevented from generating information about particular atrocities, thus limiting or preventing memorial efforts of communities who wish to remember these events publicly. Authoritarian-leaning states seeking to limit or censure memorialization activities may use AI-related technologies to establish a preferred narrative of the past while stifling expressions that stray from that norm.[7]

Alternatively, AI can potentially produce watered-down or even irrelevant generalities about a given atrocity at the expense of more nuanced complexities that make memorialization meaningful, thus enabling a different form of algorithm-driven "masked censorship" [57, p. 38]. In some cases, the motivation for censorship can be benign - i.e., in the case of commercial companies behind AIs forbidding the generation of atrocities-related content to avoid potential misuse of AIs for trivialization and denialism - but it nevertheless can interfere with the use of technology for memorialization.

- *AI can be used to generate narratives that support distortion and denialism.*

Similar to the possibilities for amplifying genuine memory practices, generative AIs can be abused to amplify distortion and denial in the context of mass atrocities. Such abuses can be intentional and unintentional. In the former case, AIs can be used to produce a large volume of diverse content promoting the claims denying that specific instances of mass atrocities did not happen (e.g., via text-focused AIs) or promoting the distorted representation of the atrocities (e.g., via image-focused AIs generating sexualized images of Holocaust victims, such as Anne Frank, or images glorifying the perpetrators). Distorted representations could even be quite nuanced, for example, by producing generative images of atrocities that look similar to the authentic ones but lacking an important detail or replacing the faces of perpetrators to confuse the public.

Aside from the risks of intentional abuse of the technologies, generative AIs are prone to hallucinations [5], in which the systems fill up information voids with generated content that is not factually supported. An example of such hallucination in the case of mass atrocities can be the invention of fictional details about a particular perpetrator or a victim to respond to the user prompt in response to which the AI has no specific information. In this case, the risk is intrinsic to the technology itself (i.e., the ability to generate content): without establishing a system of controls (e.g., allowing the user that there is no information available about their inquiry), the use of AI can lead to the distortion of historical facts.

---

[7] Some examples include several countries in Southeast Asia [46,74,79], China [15,16], Russia [50], Turkey [24].

- *AI can undermine trust in the use of technology for memorializing mass atrocities*

The humanlike responses of AI chat systems to users' questions make these systems inherently more trustworthy to the humans who engage with them [27]. Science fiction films like Spike Jonze's *Her* (2013) and Alex Garland's *Ex Machina* (2014) have contributed to the popular perception of AI as compelling – but also potentially dangerous and untrustworthy. The inherent uncertainty that accompanies trust [67] can be integral to engaging in new experiences and, in some cases, goes so far as to be a 'leap of faith'[8] [99] - for example, *in extremis* in Garland's film or potentially for some of today's generative AI users. With respect to the transmission of genocide and other mass atrocity events, however, the impacts of a potential 'leap of faith' by users engaging with AI chat systems could have negative results. For example, users may fail to check questionable claims generated on chat systems and potentially repeat them.

The unsuccessful leaps of faith in relation to the use of generative AI for f memorialization can interfere with the successful integration of these technologies in memorialization practices, as such integration is predicated (as it is for other tasks [22]) on a degree of trust in the technology. There needs to be some level of certainty that the AI system is conforming to expectations both in its use and its output. When applied to subjects as delicate as mass atrocity memorialization, accuracy is often viewed as a key imperative. The guiding transnational belief is that the atrocities should be remembered, and the suffering of victims should be honored and not forgotten [19], and for this purpose to be achieved, the accuracy of representation of atrocities and suffering must be trusted. With this in mind, if a specific form of generative AI were to acquire a reputation for inventing its facts or promoting skewed interpretations of mass atrocities, an erosion of that underlying trust would likely occur. Such AI-based foibles could lead to an overall distrust of new technology.

Given that generative AI's output is (at least partially) determined by the training data selected by its creators, critics of the technology sector might seek to disparage AI based on the trope that "big tech" routinely censors knowledge production in the interests of a hidden agenda – which, per the logic of conspiracy theory, need not even be specified to be believed. Moreover, irrespective of the systems themselves, individuals may still distrust the AI-produced text and narratives due to unverified claims made about the content's veracity. Under these circumstances, the malperformance of certain forms of generative AIs could create a sort of moral panic, causing severe distrust towards new forms of innovation similar to what had happened when earlier non-generative AI systems spawned fears of filter bubbles [9], and led to distrust towards their use in societally relevant sectors, such as journalism.

- *AI can disclose sensitive information about individuals*

---

[8] Anthony Giddens [25] notes that trust involves risk, and therefore a 'leap of faith'.

Generative AI models are trained on vast amounts of training data collected from the Internet, and such data may include personal information. Since 2021, research has demonstrated that it is possible to extract this personal information from earlier AI models [13], and ChatGPT is not exempt from this issue [49]. Similarly, it is also possible to retrieve training images (or sections of them) from models that generate visual content [14]. The sudden launch of these models to the market has left governments scrambling to take action against generative AI providers, for instance, in the case of Italian data protection regulators demanding OpenAI stop using training data of Italian citizens as it is not compliant with the General Data Protection Regulation (GDPR). In response, OpenAI has stopped providing access to chatGPT in Italy [11].

While the disclosure of sensitive information per se is problematic, in the case of the memorialization of mass atrocities, it raises additional concerns due to the care and sensitivity integral for respecting and honoring victims. Sensitive details about the lives of victims of mass atrocities, including, for instance, images of them being murdered or tortured, prisoner intake photographs, or other visual representations (including the ones taken before the atrocities, such as victims' childhood photos) may be problematic in the context of the production of new types of content, for public (or even commercial[9]) purposes. However, the exclusion of this material is challenging given the volume of content available as well as the difficulty of drawing boundaries regarding such exclusion.

- AI can create non-authentic content in relation to mass atrocities

The ability of AI to generate content that can be hardly differentiated from the content produced by a human (e.g., [84]) raises concerns that generative AIs could produce inauthentic content about a mass atrocity which might then be presented and treated as authentic. Such content can vary from fabricated historical documents assigning the blame to specific actors or whitewashing the actual perpetrators to fake visual evidence offering a distorted representation of the atrocities (e.g., to scapegoat a specific group of victims or bystanders). While companies behind AIs will likely try to limit such uses, the susceptibility of models to blindly follow user instructions can result in users circumventing these restrictions in a process known as "jail-breaking" as demonstrated in studies examining the potential of generative AIs to produce extremist content (e.g., [64]). If people believe potentially false statements made by an AI, they can fail to check the results against other sources, which can lead to perpetuating false claims and exacerbate the problems of denial and distortion already present. Even if users of generative AIs remain capable of detecting and rejecting misleading or extremist claims, there could still be a cumulative effect of undermining the trust in historical sources through the generation of more subtle content that steers the narrative toward interests beyond the genuine purpose of atrocity memorialization itself.[10]

**Other considerations of using generative AI for digital memorialization of mass atrocities**

---

[9] See [98] for the discussion of how photographs of prisoners at Tuol Sleng were digitally altered for commercial wall art and sold on major shopping sites.
[10] This idea stems from the observation by [88] concerning that exposure to AI-generated deepfakes can erode trust in social media in general.

Beyond the possibilities and potential risks identified in the previous sections, there are additional considerations and dimensions related to applying generative AI technologies to mass atrocity memorialization practices. Some of these considerations emanate from challenges that may arise in research efforts to understand the impacts of AI on memorialization, whereas others relate to the long-term implications of integrating generative AIs in the broad field of digital memorialization.

- *Access to data about mass atrocities*. The widely used AI models usually rely on information available on the Internet, but most of the content related to mass atrocities is stored in traditional archives, which are often only partially digitized. Therefore, the knowledge that AIs possess about atrocities is limited to already digitized or digital-born content (e.g., PDFs, images, and other content items, including the ones coming from non-institutional sites such as Pinterest), which the model can access. Another aspect of data access concerns the tendency of some generative AIs (e.g., chatGPT-3) to rely on the snapshot of data produced at some time point (e.g., only up to 2021). Time-range boundaries can have implications for content generation in relation to mass atrocities, especially concerning the atrocities which happened after the model underlying the AI has been trained and deployed. At the same time, many AI models integrate possibilities for retrieving data from the Web to overcome these time limits; however, it also increases the possibility of using unverified or false data.
- *Matters of user privacy*. Currently, there is limited understanding of how generative AI platforms use the data provided by their users (e.g., whether they track information requests generated by individual users) and what applications of these data they might consider. In the case of data about mass atrocities, certain requests might be rather sensitive (for instance, inquiries about the war in Ukraine contradicting the official state narrative in Russia for Russian users, in particular when using Russia-based AI platforms).
- *Matters of representation*. With respect to the Let Them Speak project, Naron and Toth [69] assert that the aim of the project is to make the fragmented collection of Holocaust testimonies representative of those who perished and, therefore, could not speak. Applying generative AI could take such projects a step further by using separate testimonies to give AI the power to speak and tell coherent stories about victims' experiences. On the positive side, such a system arguably permits a representation of those who have no voices anymore. However, this move begins to approach the uncanny valley where the chatbot speaker could become a victim of a sort itself or, interpreted through a supernatural lens, as a medium for the ghosts and spirits of the deceased. A chatbot could conceivably become seen as an entity unto itself, distinct from the inputs used to construct its persona (and memories, of a sort). Representing those who have perished in a mass atrocity through a technology that uses existing testimonies to fabricate new ones may also raise ethical concerns about atrocity remembrance along with epistemological ones about how we understand and represent the past.
- *Replacement of the human working force involved in memorialization*. The ability of generative AI to produce a large volume of textual/visual content related to mass atrocities enables new possibilities for decreasing the costs of digital memorialization activities. However, these possibilities also create risks for humans working in memorialization-related fields. For example, visual artists, copywriters, data analysts, and certain types of archivists could potentially be reduced or replaced by generative AI technologies. Additionally, generative AIs can perform some functions associated with the roles of curators working in heritage institutions connected with mass atrocities. Not only shall we consider the economic implications of job loss or displacement, but the fact that the involved individuals, directly or indirectly, increase awareness of mass atrocities and propagate the sensitivities of the past suffering to groups that would otherwise not engage with it. Under these circumstances, it is imperative to consider integrating generative AIs into memorialization practices while prioritizing the input and expertise professionals in these sectors have and the importance of social interaction in informing individuals in other societal sectors.

**An agenda for future research on the role of generative AI in the context of atrocity memorialization**

In light of the trends, potentials, and uncertainties in AI-generated atrocity memorialization that we have identified, we now turn to the question of what we, as a scholarly community, can do to realize the possibilities and counter the risks of using generative AI for digital memorialization. Below, we share four recommendations that, in our view, are integral for steering the use of these technologies toward a more ethical and sustainable direction.

*Experimenting with AI in the context of mass atrocities*. We recognize threats related to genocide distortion/denialism, which can be amplified by AI, but we should not ignore the possibilities provided for the memorialization of mass atrocities. To examine these possibilities, we need to keep experimenting with generative AI and also keep options for experimentation open for other stakeholders, including ordinary citizens. The ability to experiment with AI is essential for understanding the full range of its possible uses for memorialization. While banning the atrocities-related prompts (as done by certain AIs, such as Bing Image Creator) can be a safe option, it could also, problematically, contribute to erasing memories about past suffering by undermining memorialization (and prevention and rescue) efforts. For example, human rights groups seeking to collect and display evidence of war crimes and genocide may have the content of their projects censored out from AI training data (and, consequently, AI system outputs), thus making it difficult for them to show the crimes being committed with the hope of some action [37, p. 10]. Several studies [23,37,38] have observed a similar phenomenon concerning human and algorithmic moderation systems used by social media platforms to block potentially offensive content (including evidence of mass atrocities) then delete such content before perpetrators can be held accountable. Finally, even if major AI companies ban queries dealing with mass atrocities, it might be hard to do it comprehensively (e.g., so users would not find a way to circumvent the limitations), whereas such bans can push individuals towards using generative AI models deployed in fringe environments like the dark web.

*Studying how generative AI is used for preserving and distorting memories of mass atrocities in and outside academia*. There are indications (e.g., [80]) that generative AIs could become another source of moral panic, similar to earlier forms of AI-driven technology (e.g., recommender systems and the associated filter bubble fears [9]). To counter it, it is important to understand how generative AIs are used in the memorialization of mass atrocities and whether the concerns about the risks of their adoption in the field of memorialization are justified. The complexity of memorialization, which includes a broad range of individual and

group practices focused both on remembering past suffering and on researching such practices, adds to the challenging nature of this task.

How to achieve such an understanding is an open question. Unlike content released via websites, blog posts, or social media, memorial activity relying on AI platforms is often confined to private accounts. The inputs and outputs of AI systems are hidden from public view unless or until the user shares them in a public (presumably digital) forum. While there are exceptions (e.g., the rendering of images on Discord servers using models like Midjourney), even in these cases, the degree of communal engagement with generative AI appears to be more limited than with more conventional digital memorialization practices such as commentaries for statements of memorial institutions published on social media.

Under these circumstances, a first step can be to examine cases where individuals or groups integrate AI-rendered text, image, video, or voice content into established memorialization practices (e.g., by adding them to memorial websites devoted to a mass atrocity) and, where possible, conduct interviews with practitioners involved in such integration. Though few currently, such cases are likely to proliferate as individuals and groups seek the capabilities that AI can provide to enhance the message they want to convey about past suffering. Another alternative can be to use experimental research designs to examine how individuals use AIs to generate memorialization-related content or engage with such content in a controlled environment. Similarly, one might track the extent to which heritage institutions and human rights organizations adopt generative AIs, along with their respective motivations and expectations.

*Arguing for the importance of integrating transparency principles in the generative AI design and functionality*. It is essential that generative AI design follows the principles of transparency both in data acquisition for training the underlying models (e.g., to ensure that data was acquired without breaching intellectual rights) and in providing possibilities for analyzing the performance of generative AIs and their potential bias. Similarly, greater transparency is needed in the functionality of generative AIs: for instance, by providing sources for data that are used to generate outputs (when possible) so potential privacy or copyright breaches would be easy to identify. Transparency is especially important in the context of using generative AIs for producing content related to mass atrocities due to associated ethical and moral considerations (e.g., the importance of treating the memory of victims respectfully) as well as the need for public cooperation in the effort to identify potential risks of AI misuse. Under these circumstances, transparency of AI design and functionality becomes a main prerequisite for AIs becoming reliably truthful and establishing a reputation as such.

*Watch AI developments beyond the West and Silicon Valley*. The development and application of generative AIs to mass atrocity memorial practices are not confined to the West and Silicon Valley. Generative AIs are employed, refined, and redesigned to meet the needs and tastes of various cultures and nationalities around the globe. For instance, the Cambodian government is currently creating its chatGPT system for the public [51], and another version was created in Saudi Arabia (miniGPT-4 [61]). While it is not clear yet when or how these platforms will be used for memorial purposes, there is little question they will eventually be adopted to memorialize past and present suffering. In China, for example, the government has sought to place guardrails on the range of sources a chatbot might use to discuss particular events, such as the Tiananmen Square massacres [16], and the Cambodian government is evaluating the risks of the public using the Khmer chatGPT as a primary source of information [94]. Moreover, the speed at which these technologies are being developed worldwide is astounding. Cambodia, which is considered a developing nation and has been struggling for decades to recover from the devastating Khmer Rouge era, has already produced a robot that speaks using the model underlying the ChatGPT platform [75]. What do these developments suggest for memorialization? How will these varying forms of ChatGPT impact the memorial narratives produced? We can expect the unexpected: developments that Silicon Valley could not have anticipated as the AI models are increasingly trained implicitly and explicitly based on different core ideologies, cultures, and political agendas.

**Conclusion**

The foregoing insights about the impact of generative AI on mass atrocity memorialization must come, as is the case for any future-looking discussion on AI, with the caveat that there are more unknowns than there are knowns. As the developments in the field of AI accelerate, academic research and societal discussions must remain adaptable and responsive to changing developments. This article offers some guidelines in terms of what scholars, as well as practitioners in both the AI and memorialization fields, may need to consider at this moment. As is the case with other rapidly evolving subjects, the discussion of possibilities and risks of relying on generative AI in the context of memorialization, as well as considerations associated with such reliance, can help to establish a framework of reference for anticipating practical and conceptual issues and guiding future debates on the topic. Transparent and collaborative efforts drawing from various academic disciplines can help to address the challenges presented by this emergent technology to support its ethical integration with existing memorialization practices.

Here: